# APPLYING ADVANCED SPACEBORNE THERMAL EMISSION AND REFLECTION RADIOMETER (ASTER) SPECTRAL INDICES FOR GEOLOGICAL MAPPING AND MINERAL IDENTIFICATION ON THE TIBETAN PLATEAU

R. K. Corrie [a, *], Y. Ninomiya [b], J. C. Aitchison [a]

[a] Tibet Research Group, Department of Earth Sciences, University of Hong Kong, Hong Kong SAR, China
[b] Geological Survey of Japan, AIST, Tsukuba, Japan

**ABSTRACT:**

The Tibetan Plateau holds clues to understanding the dynamics and mechanisms associated with continental growth. Part of the region is characterized by zones of ophiolitic mélange believed to represent the remnants of ancient oceanic crust and underlying upper mantle emplaced during oceanic closures. However, due to the remoteness of the region and the inhospitable terrain many areas have not received detailed investigation. Increased spatial and spectral resolution of satellite sensors have made it possible to map in greater detail the mineralogy and lithology than in the past. Recent work by Yoshiki Ninomiya of the Geological Survey of Japan has pioneered the use of several spectral indices for the mapping of quartzose, carbonate, and silicate rocks using Advanced Spaceborne Thermal Emission and Reflection Radiometer (ASTER) thermal infrared (TIR) data. In this study, ASTER TIR indices have been applied to a region in western-central Tibet for the purposes of assessing their effectiveness for differentiating ophiolites and other lithologies. The results agree well with existing geological maps and other published data. The study area was chosen due to its diverse range of rock types, including an ophiolitic mélange, associated with the Bangong-Nujiang suture (BNS) that crops out on the northern shores of Lagkor Tso and Dong Tso ('Tso' is Tibetan for lake). The techniques highlighted in this paper could be applied to other geographical regions where similar geological questions need to be resolved. The results of this study aim to show the utility of ASTER TIR imagery for geological mapping in semi-arid and sparsely vegetated areas on the Tibetan Plateau.

## 1. INTRODUCTION

### 1.1 Background

The Tibetan Plateau is often referred to as the "roof of the world", and contains some of the highest topography on the planet, with an average elevation exceeding 4,500 m. The region holds special geological significance, owing its existence to the continent-continent collision between India and Asia. It thus offers an opportunity to understand tectonic processes in addition to the mechanisms and timing for the India-Asia collision.

The Tibetan Plateau is composed of a number of E-W oriented continental fragments, or terranes, accreted over time onto the southern margin of the Eurasian plate. A series of suture zones separate the terranes, with the Yarlung Tsangpo Suture (YTS) and the Bangong-Nujiang Suture (BNS) forming boundaries to major tectonic units in Tibet. Ophiolitic mélanges are associated with the suture zones and are thought to represent ancient remnants of oceanic lithosphere and underlying upper mantle, formed at spreading centres and subsequently emplaced during oceanic closures. They preserve clues to their evolution and the timing of continental accretion and growth. Many of the ophiolitic outcrops are located in extremely remote and isolated areas of Tibet and access is limited due to the difficult field conditions and inhospitable terrain. Thus only a few areas on the Tibetan Plateau have been mapped in any sufficient detail and remote sensing remains the only technique for making a more detailed investigation.

The last ten years have witnessed a new generation of imaging satellites capable of discriminating rock types at a scale useful for geological applications. One of these satellites – the Advanced Spaceborne Thermal Emission and Reflection Radiometer (ASTER), offers improvements over previous satellites in areas of spatial, spectral, and temporal resolution. Previous efforts to map ophiolites using multispectral remote sensing include studies using the Landsat series of satellites, where the Landsat Thematic Mapper (TM) has been widely used in Egypt (Sultan et al., 1986, 1987), the Oman (Rothery, 1987a, 1987b; Abrams et al., 1988), Alaska (Harding et al., 1989), Tibet (Matthews and Jones, 1992), and along the Indo-Myanmar border (Jha et al., 1993). More recently, the Indian Remote Sensing Satellite (IRS-1C/1D) has been used in the Himalaya, India (Philip et al., 2003), whilst ASTER imagery has been used in Tibet (Ninomiya, 2003; Ninomiya et al., 2005) and Pakistan (Khan et al., 2007; Khan and Mahmood, 2008).

Spectral mapping of different rock types requires advanced digital image processing techniques. Techniques used in ophiolite discrimination have included band ratios (Sultan et al., 1987; Matthews and Jones, 1992; Ninomiya et al., 2005), decorrelation stretching (Abrams et al., 1988; Matthews and Jones, 1992; Philip et al., 2003), false colour composites (Harding et al., 1989), IHS transformation (Philip et al., 2003), principal components analysis (Khan et al., 2007), minimum noise fraction (Khan et al., 2007), and log residuals (Khan and Mahmood, 2008). However, mapping ophiolites based on their spectral characteristics presents challenges. This is because ophiolites do not contain a single rock type, but are representative of a wide range of lithologies. Ninomiya of the Geological Survey of Japan has pioneered the use of several mineralogical indices using ASTER thermal infrared data (Ninomiya et al., 2005).

This paper assesses the utility of ASTER thermal infrared band ratios for the purposes of evaluating their effectiveness in the discrimination and mapping of ophiolites and other surface lithologies. A region in western-central Tibet was selected as a test area, and the results were compared with published maps and from ground based *in-situ* observations from the field.

---

* Corresponding author. E-mail address: corrie@cantab.net

## 1.2 Geological setting

The study area is located to the east of the town of Gertse in the western sector of the BNS, western-central Tibet (Figure 1). The region is dominated by E-W trending strike slip and thrust faults, and contains the three major lakes: Lagkor Tso (32°02'N, 84°07'E), Dong Tso (32°16'N, 84°40'E), and Zhaxi Tso (32°12'N, 85°06'E). The regional geology includes a diverse rock association including Triassic-Jurassic limestone, Cretaceous volcanic rocks and non-marine strata, Jurassic flysch, Quaternary deposits and outcrops of ophiolitic mélange (Wang, 2008).

Two occurrences of ophiolitic rocks are present in the study area. These are located to the north of Lagkor Tso and to the north and east of Dong Tso (identified in black in Figure 1). The Lagkor Tso ophiolite consists of a disrupted but mostly complete sequence, dominated by a serpentinite-matrix mélange, and includes metaperidotites, layered gabbros, basaltic pillow lavas, and metamorphic rocks. The Dong Tso ophiolite preserves a complete ophiolitic suite of rocks, which has been somewhat disrupted through extensive faulting, and includes harzburgitic peridotites, serpentinites, layered gabbros, a sheeted dyke complex, and pillow basalts (Wang, 2008).

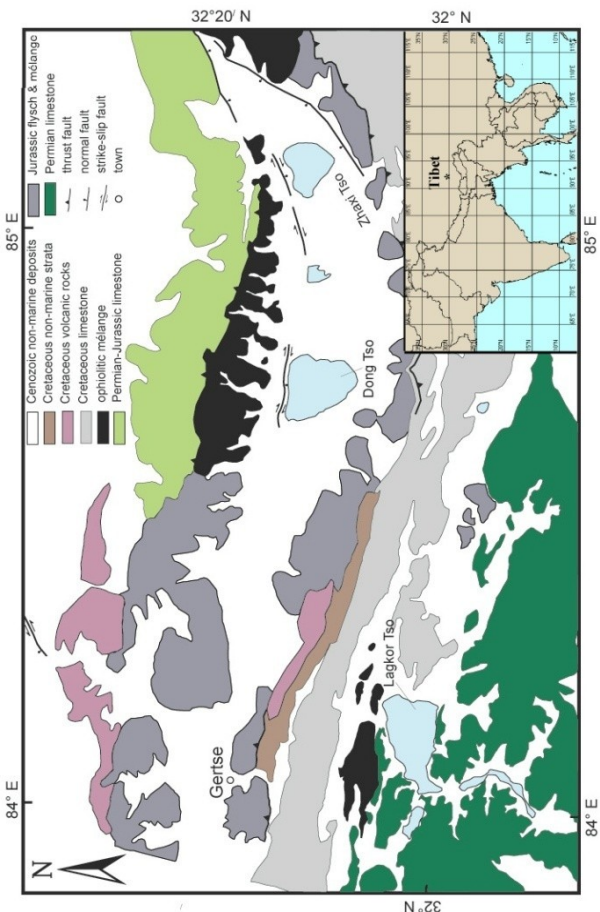

Figure 1. Simplified geological map of the Gertse area of western-central Tibet. After Wang (2008).

## 1.3 An overview of ASTER

The ASTER instrument (Fujisada, 1995; Yamaguchi et al., 1998), launched in December 1999, aboard the Terra satellite, is a cooperative effort between NASA, Japan's Earth Remote Sensing Data Analysis Centre (ERSDAC), and Japan's Ministry of International Trade and Industry (MITI). ASTER offers medium to high spatial resolution, a ground swath width of 60 km, and a revisit time of 16 days. The instrument is composed of three optical imaging subsystems acquiring fourteen multispectral bands operating in the visible near infrared (VNIR), shortwave infrared (SWIR), and thermal infrared (TIR) channels, each achieving spatial resolutions of 15 m, 30 m, and 90 m respectively.

ASTER offers substantial improvements over previous earth observation satellites such as the Landsat Thematic Mapper (TM), and Enhanced Thematic Mapper (ETM+) series. One improvement is seen in its thermal infrared capability, where ASTER offers 5 channels at a spatial resolution of 90 m (Table 1) - compared to TM and ETM+, which provide 1 channel at 120 m and 60 m respectively.

| Sub-system | Band number | Spectral range (µm) | Centre λ (µm) | Spatial resolution |
|---|---|---|---|---|
| TIR | 10 | 8.125 – 8.475 | 8.3 | 90 m |
| | 11 | 8.475 – 8.825 | 8.65 | |
| | 12 | 8.925 – 9.275 | 9.1 | |
| | 13 | 10.25 – 10.95 | 10.6 | |
| | 14 | 10.95 – 11.65 | 11.3 | |
| Ground swath width = 60 km | | | | |

Table 1. ASTER TIR instrument characteristics.

## 1.4 ASTER TIR indices defined

Ratio images (or band ratios) divide the digital number (DN) values of one band by the corresponding DN values of another band (or some other physical quantity, such as spectral radiance), displaying the result as a grayscale image. If ratios of bands are selected targeting specific spectral absorption features, then the resultant image can help to identify the spectral variability and composition within a scene.

Based on an analysis of TIR spectral characteristics of rocks, Yoshiki Ninomiya of the Geological Survey of Japan has pioneered the use of several spectral indices for the mapping of quartzose, carbonate, and silicate rocks using ASTER data. Ninomiya's indices are summarized in the remainder of this section.

The carbonate minerals, calcite and dolomite, have a unique spectral absorption feature in ASTER band 14 (Figure 2a). As a result, based on differences in spectral emissivity between bands 13 and 14, the Carbonate Index (CI) is defined as:

$$CI = \frac{Band_{13}}{Band_{14}} \quad (1)$$

CI is expected to be high for the carbonate minerals, calcite and dolomite, but is not expected to identify other carbonate rocks.

Pure silica ($SiO_2$) bearing rocks (quartzose), such as quartzite, display a characteristic spectral feature between ASTER bands 10 to 12 in the thermal infrared. It is found that emissivity is higher at ASTER band 11 than at bands 10 and 12 (Figure 2b). Based on these spectral differences, the Quartz Index (QI) is defined as:

$$QI = \frac{Band_{11} \times Band_{11}}{Band_{10} \times Band_{12}} \quad (2)$$

QI is expected to be high for quartzite and other siliceous rocks, but low for potassium feldspar and gypsum.

Silicate minerals and rocks have unique spectral properties in the thermal infrared. It is observed that the emissivity minimum moves to longer wavelengths as the chemical $SiO_2$ content of the rock type decreases (represented by arrow in Figures 2b-f). Secondly, more felsic rock types (i.e. granite) show a higher emissivity at longer wavelengths (i.e., band 13), whereas more ultramafic types (i.e. peridotite) show a higher emissivity at shorter wavelengths (i.e., band 12). These spectral properties can be exploited to define the Mafic Index (MI):

$$MI = \frac{Band_{12}}{Band_{13}} \qquad (3)$$

It is expected that MI will increase as the ratio of the two bands moves from a felsic to an ultramafic rock type. This is because MI has an inverse correlation to the $SiO_2$ content in silicate rocks - as MI increases, the silica content decreases, and vice versa.

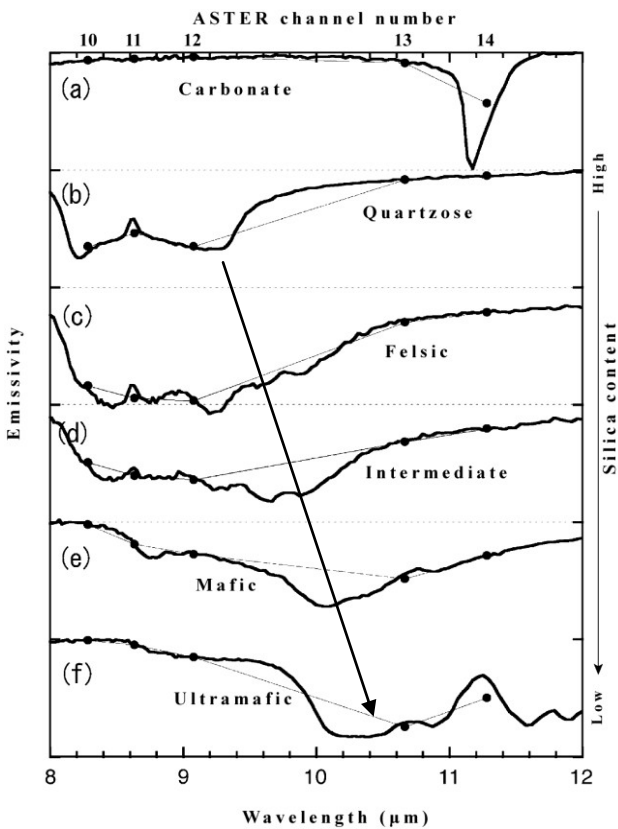

Figure 2. Emissivity spectra for common rock types.
After Ninomiya et al. (2005).

In addition to being sensitive to the $SiO_2$ content in silicate rocks, it is found that MI is also sensitive to carbonates. To reduce this unwanted effect, Ninomiya introduced a carbonate separation of n = 3 into his mafic index (Ninomiya et al., 2005):

$$MI(n) = \frac{Band_{12}}{Band_{13} \times CI^n} = \frac{Band_{12} \times (Band_{14})^n}{(Band_{13})^{n+1}} \qquad (4)$$

$$MI(3) = \frac{Band_{12} \times Band_{14} \times Band_{14} \times Band_{14}}{Band_{13} \times Band_{13} \times Band_{13} \times Band_{13}} \qquad (5)$$

where, n = degree of $CaCO_3$ separation; CI = Carbonate Index
Equation (3) represents an MI separation of zero, i.e., MI(0).

Table 2 shows various igneous rock categories with examples of each rock type, relative $SiO_2$ content, and emissivity profile. For a more in-depth discussion of QI, CI, and MI, the reader is directed to Ninomiya (2004) and Ninomiya et al. (2005).

| Rock category | Example rock type | $SiO_2$ content | Emissivity in band 12 / 13 |
|---|---|---|---|
| Felsic | Granite | High | Low / High |
| Intermediate | Diorite | ↓ | Low / High |
| Mafic | Gabbro |  | High / Low |
| Ultramafic | Peridotite | Low | High / Low |

Table 2. Rock types and $SiO_2$ content.

## 2. METHODOLOGY AND IMAGE PROCESSING

Eight ASTER Level 1B (L-1B) scenes (recorded in the winter months between 2000 and 2004) were acquired from Japan's Earth Remote Sensing Data Analysis Centre (ERSDAC, http://www.ersdac.or.jp). The ASTER data was delivered in a Hierarchical Data Format (HDF) which provides two files for each scene - a binary .hdf file containing the imagery and an ASCII text .met file containing the metadata. The imagery was checked for acceptable levels of cloud cover (< 2%) and for sensor errors, such as banding and other geometric distortions.

The TIR channels (bands 10-14) were extracted, and the raw Digital Number (DN) values converted to physical units of spectral radiance using the Unit Conversion Coefficients (UCCs). The UCCs, supplied in the .met file, correct for sensor gain and offset and are different for each ASTER band and for each gain setting used to acquire the image. ASTER L-1B imagery is supplied in terms of scaled radiance at-sensor data with radiometric and geometric corrections applied.

At-sensor spectral radiance (units of $W\ m^{-2}\ sr^{-1}\ \mu m^{-1}$) may be calculated for each DN value of an ASTER scene using the following expression (Abrams et al., 1999):

$$L_{sen_i} = (DN_i - 1) \times UCC_i \qquad (6)$$

where, $L_{sen_i}$ = spectral radiance in $band_i$ ($W\ m^{-2}\ sr^{-1}\ \mu m^{-1}$)
$DN_i$ = Digital Number in $band_i$
$UCC_i$ = Unit Conversion Coefficient in $band_i$

The amount of radiation from a surface reaching the satellite sensor is a function of its temperature, emissivity (composition) and the effects of the atmosphere. However, no atmospheric correction was applied to the imagery. The indices seem stable against variations in temperature and atmospheric conditions, and previous work shows a good correlation between ASTER L-1B TIR imagery and surface geology (Ninomiya et al., 2005).

Water features, snow, and clouds were removed from the imagery by setting their radiance values equal to zero. This was achieved through an initial supervised parallelepiped classification of the VNIR and TIR bands and applying the product of the classification to the TIR channels as a binary mask.

The indices, QI and MI, appear unaffected by surface temperature variations within a scene if atmospheric conditions are good. However, it is observed that CI is directly affected by surface temperature changes (Ninomiya, 2002). This is because L-1B data is supplied without radiometric correction. To eliminate the temperature component all ASTER L-1B at-sensor radiance values were normalized by converting the brightness temperature of ASTER TIR $band_{13}$ to a fixed value of 300 K using Planck's law. This creates a simple form of temperature emissivity separation by reducing the dependency of CI to surface temperature (Ninomiya, 2002; Ninomiya and Fu, 2003; Ninomiya et al., 2005). Once normalized, the results suggest a marked improvement in the ability of CI to detect carbonate rocks. Normalized ASTER L-1B data can be calculated from Planck's law:

$$\hat{L}_{sen_i} = L_{sen_i} \times \frac{exp\left\{\frac{\lambda_{13}}{\lambda_i}\ln\left(\frac{c_1}{\pi(\lambda_{13})^5 L_{sen_{13}}} + 1\right)\right\} - 1}{exp\left(\frac{c_2}{\lambda_i\, nT/\epsilon_{13}}\right) - 1} \quad (7)$$

where, $L_{sen_i}$ = spectral radiance in $band_i$ (Wm$^{-2}$sr$^{-1}$μm$^{-1}$)
$\lambda_i$ = centre wavelength of $band_i$ (μm)
$c_1$ = constant 1 = $3.74 \times 10^8$ W m$^{-2}$ μm$^4$
$c_2$ = constant 2 = $1.44 \times 10^4$ μm K
$nT$ = temperature (K) = 300 K
$\epsilon_{13}$ = emissivity in $band_{13}$ = 1.0

The emissivity in $band_{13}$ is assumed to be 1 (blackbody); the radiation constants $c_1$ and $c_2$ are equal to $2hc^2$ and $hc/k$ respectively; h is Planck's constant (6.626 x 10$^{-34}$ J s); c is the speed of light in a vacuum (2.998 x 10$^8$ m s$^{-1}$); and k is Boltzmann's constant (1.38 x 10$^{-23}$ W s K$^{-1}$). The centre wavelengths of the TIR bands can be obtained from Table 1.

QI, CI, and MI were applied to normalized at-sensor radiance imagery producing grayscale images for each satellite scene. A false colour composite image (FCC) was then created for each scene by sending the indices to the separate colour guns of the display. Finally, the FCC images were mosaicked and a linear contrast stretch applied before final output. It is important to note that no colour correction was used during the mosaicking process. This was considered a more favourable approach compared to mosaicking the data set first and applying indices second. The reasoning for this is that each satellite scene contains scene dependent spectral radiance values which vary according to date and time of day of acquisition and are unique to each image. Mosaicking and radiometrically normalizing TIR satellite data are discussed further by Scheidt et al (2008). To create a consistent workflow all indices were applied to normalized L-1B imagery.

## 3. DISCUSSION OF RESULTS

Processed grayscale images of QI, CI, and MI are shown in Figures 3A, B, and C respectively. The images were enhanced with linear contract stretches of 1.0 to 1.05 for QI, 1.01 to 1.045 for CI, and 0.80 to 0.90 for MI. Limits were set for detecting pixels higher than a given threshold. Detected pixels, where QI ≥ 1.055 (representing a high quartz signal), are indicated in red (Figure 3A). A second threshold was set for CI ≥ 1.05 (high carbonate signal), with detected pixels shown in green (Figure 3B). Finally, a third threshold was set, where MI ≥ 0.94 (high mafic signal), with detected pixels highlighted in purple (Figure 3C). The grayscale images used in combination with the FCC image can help in the geological interpretation of the satellite scene. While the grayscale images are useful for identifying individual rock types, the FCC image is superior for understanding the spatial extent of the geological outcrops. The use of both image types is essential when making a detailed geological analysis of the study area.

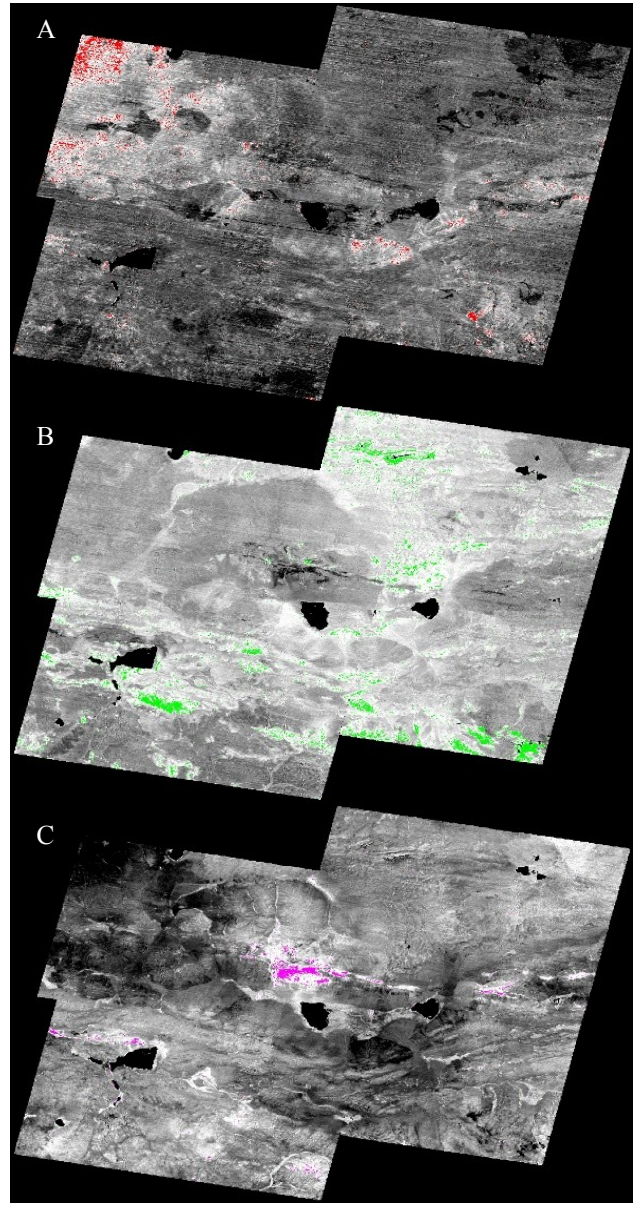

Figure 3. Grayscale images of QI, CI, and MI. Coloured regions identify pixels above threshold values.

Figure 4 shows the FCC image of the ASTER TIR data after assigning QI to the red, CI to the green, and MI to the blue colour guns of the display. The three components have had linear contrast stretches applied to their histograms of 0.97 to 1.055 (R), 1.005 to 1.055 (G), and 0.79 to 0.95 (B) for enhancement purposes. Lakes and other water features have been masked out to black. Generally, an orange-red colour in the FCC image represents quartz-rich, feldspar-poor (i.e. mostly quartzose sedimentary rocks), a light green colour represents areas rich in carbonates, i.e. calcite and dolomite-rich limestone, and a blue-purple colour identifies silicate rocks. Ophiolites are composed of mafic - ultramafic rock types (low $SiO_2$ content) and are expected to present as a purple colour in the FCC image.

Results from the image processing agree well with existing geological maps and recorded field data. However, the study area had not been mapped in any degree of detail. Limited geological maps were available for only part of the study area and detailed maps were not available at all. For this reason, ground truthing will need to be carried out at a later date.

Identifying labels are presented on the FCC image showing the locations of features that are further discussed in the text that follows. The large homogenous region located in the NW corner of the FCC image (identified in orange and labelled as "A" in Figure 4) has high QI, low to intermediate CI, and low MI (see corresponding images in Figure 3). From its high QI values, region "A" is expected to be quartz-rich, feldspar-poor sedimentary rocks and / or deposits. This is confirmed from the geological map, where Jurassic flysch and mélange in the NW and central region is indicated. Flysch is quartz-rich and would explain the high QI signal. Similar quartz-rich areas can also be observed south of Dong Tso and Zhaxi Tso.

Region "B" is expected to be predominantly carbonate rock, based on its high CI values (displayed as light green in Figure 4). Light green areas in the image show intermediate QI, high CI, and intermediate MI. CI is expected to be high for those areas rich in the sedimentary rock limestone (sensitive to the major carbonate minerals calcite and dolomite). However, CI is not expected to correlate with other carbonate types (Ninomiya et al., 2005). The dark green area to the west of Zhaxi Tso and to the east of Lagkor Tso, located in Quaternary deposits ("C"), are low QI, and are expected to be barium sulphate deposits.

Silicate rocks are identified in several regions in the FCC image. The two ophiolite outcrops at Lagkor Tso and Dong Tso (represented in black on the geological map in Figure 1) are clearly differentiated in purple from the surrounding geology in the FCC image ("D"). An ophiolite outcrop to the east of Zhaxi Tso is also suggested by the FCC image which is confirmed by the geological map. ASTER TIR spatial resolution appears to be too coarse (90 m) to be able to differentiate lithological sequences within the ophiolites, however, the sensor seems more than capable of mapping the extent of the outcrop. Immediately north of the ophiolite at Dong Tso can be found a sequence of pillow basalts that are associated with Permian limestone. The pillow basalts are easily identified in the FCC image ("E"), where they can be seen as a horizontal band running north of Dong Tso eastward towards Zhaxi Tso. Field evidence suggests that they are not a part of the ophiolite due to having undergone metamorphism (Wang, 2008). The geological map indicates the presence of igneous rocks to the north and west of the Lagkor Tso and Dong Tso ophiolites respectively. These can be identified in the FCC image ("F") and display as low QI, intermediate CI, and intermediate MI. They are mostly basaltic and andesitic in origin consisting of Cretaceous pyroclastic flows, including ignimbrites and crystalline tuff deposits (Faustino, 2009).

Areas in the FCC image displaying primary colours of red, green, and blue, represent relatively pure quartz, carbonate, or silicate content respectively and are easier to interpret. However, areas showing more complicated colours could indicate regions that are more spectrally confused and hence suggest more complicated lithologies. One such region is to the extreme south, represented by a dark green colour ("G"). It is an area of low QI, intermediate CI, and intermediate to high MI. This could represent a feldspar-rich, quartz-poor granitic body. Other regions to the south also show more complex colours that could indicate mixed materials, such as, low MI igneous rocks with marine and non-marine sedimentary rocks. A future study to ground truth these areas is expected to provide further clarification.

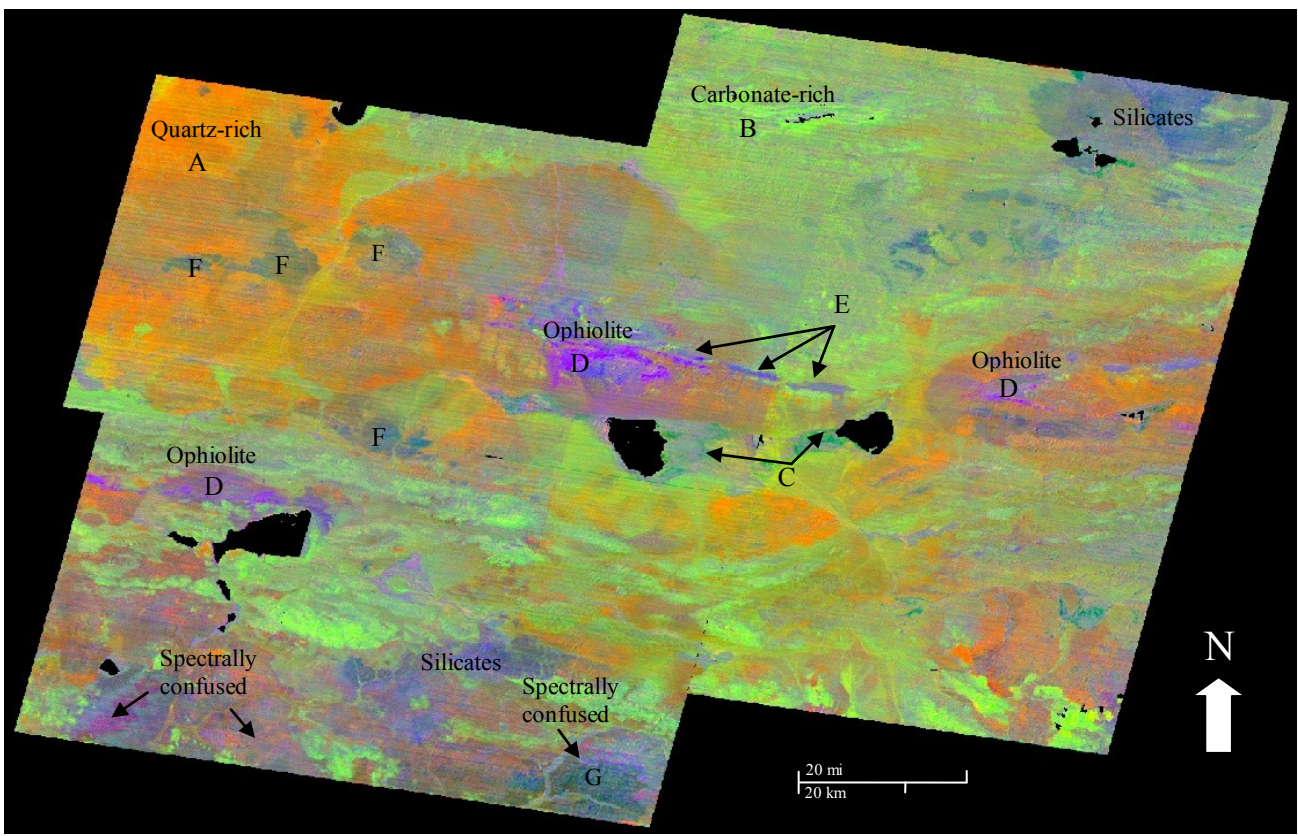

Figure 4. Colour composite image with linear contrast stretches applied to QI, CI, and MI (RGB).

## 4. CONCLUSIONS

The study area forms the western part of the ophiolite bearing BNS in western-central Tibet. ASTER TIR data were exploited using lithological indices to differentiate spectral differences in the rock units of the region. The Quartz Index (QI), the Carbonate Index (CI), and the Mafic Index (MI) were applied to multi-temporal ASTER L-1B radiance at-sensor data, without atmospheric correction. Interpretation of the imagery was completed using both gray scale and FCC images. The results agree well with the known geology and the indices prove robust in detecting quartzose, carbonate, and silicate rocks. Materials of ultramafic composition, particularly ophiolites, are clearly visible and can be distinguished from surrounding silicates. For a more complete remote sensing survey, TIR should be applied with channels from other regions of the electromagnetic spectrum, including the VNIR and the SWIR. This study shows the utility of using ASTER TIR data for geological mapping when applied to sparsely vegetated, semi- arid areas, exhibiting well exposed rocks. In theory the mapping techniques presented in this paper could be applied to other geographical regions where similar questions needed to be resolved. It is expected that this work will prepare the way for a future study with the aim of refining and automating the process so as to complete a systematic mapping of ophiolites on the greater Tibetan Plateau area.

## ACKNOWLEDGEMENTS

RKC would like to thank his colleagues Alan Baxter, Kono Lemke, Jun Liu, Doug McNeil, and Louisa Tsang of the Department of Earth Sciences, University of Hong Kong, for their review and comments in the writing of this paper. This work was supported by a research grant (to JCA) from the Research Grants Council of the Hong Kong Special Administrative Region, China (Project HKU 7001/09P).